# Un Résultat Gravimétrique pour la Renaissance de la Théorie Corpusculaire
## « An Experimental Gravimetric Result for the Revival of Corpuscular Theory »

Maurice Duval
Centre québécois de recherche et de développement de l'aluminium
637, boul. Talbot, bureau 102, Chicoutimi, Qc, Canada G7H 6A4
maurice.duval@cqrda.ca

**Abstract**
The gravitational phenomenon, founded on the assumption of absorption of a flux of gravitons through the matter, led to a law of attraction comprising a term of attenuation of gravity. The attenuation effect, which depends on the distribution of mass elements into the Sun, was compared with the relativistic effect for the orbital elements of planets. The calculations carried out with our modified law of Newton, lead to a perfect agreement for the advance of the perihelion of planets, and give an interaction cross section with the matter of $3.2 \times 10^{-17}\,\mathrm{m}^2/\mathrm{kg}$ (approximately $10^{-40}$ $\mathrm{cm}^2/\mathrm{nucleon}$). The effect of attenuation during a solar eclipse will produce a local reduction in the soli-lunar attraction of 0.13 microgal. This small variation of gravity during the eclipse can lead to distortions of the geoid of about ten millimetres of which effect on gravity – estimated at a few microgals – is compatible with the observed gravitational anomalies. The gravimetric measurements taken in the area of Montreal, during the Sun eclipse of May 10, 1994, show the existence of an anomaly of 2.4 microgals that coincides perfectly with the period of the eclipse.

-----

Le phénomène gravitationnel fondé sur l'hypothèse de l'absorption d'un flux de gravitons à travers la matière conduit à une loi d'attraction comportant un terme d'atténuation de la gravité. L'effet d'atténuation, qui dépend de la distribution de masse solaire, a été comparé à l'effet relativiste pour les éléments orbitaux des planètes. Les calculs effectués avec notre loi de Newton modifiée conduisent à un parfait accord pour l'avance du périhélie des planètes et donnent une section efficace d'interaction des gravitons avec la matière de $3.2 \times 10^{-17}\,\mathrm{m}^2/\mathrm{kg}$ (environ $10^{-40}$ $\mathrm{cm}^2/\mathrm{nucléon}$). L'effet d'atténuation pendant une éclipse solaire va produire une diminution locale de l'attraction soli-lunaire de 0.13 µgal. Cette faible variation de la gravité durant l'éclipse peut causer des distorsions du géoïde d'une dizaine de millimètres dont l'effet gravitationnel, évalué à quelques microgals, est compatible avec les anomalies gravitationnelles observées. Les mesures gravimétriques prises dans la région de Montréal, pendant l'éclipse de Soleil du 10 mai 1994, montrent l'existence d'une anomalie de 2.4 microgals en parfaite coïncidence avec la période de l'éclipse.

# 1 – Le phénomène gravitationnel et la théorie corpusculaire

La théorie de Newton de même que la théorie d'Einstein de la gravitation se fondent avant tout sur des considérations mathématiques; l'une s'appuie sur une loi d'attraction qui épouse les résultats, l'autre sur une loi de structure où la notion aristotélicienne du haut et du bas est redéfinie dans un langage mathématique moderne. Pour cette théorie moderne de la gravitation, on nous offre un Univers mathématique où la matière « déforme » l'espace et le temps. Sans réel espoir de pouvoir comprendre la nature d'une interaction entre matière et espace, et puisque le temps (notion fictive) n'est en principe rien d'autre que la relativité du mouvement, on est toujours en droit de se demander si la vérité n'est pas ailleurs.

Évitant de s'engager dans une voie spéculative, Newton souligne lui-même les limites explicatives de sa loi d'attraction en ces termes : « Que la gravitation puisse être essentielle et inhérente à la matière, qu'un corps puisse agir sur l'autre à toute distance, à travers l'espace vide, sans nul intermédiaire, de sorte que l'attraction ne soit pas conduite de proche en proche, d'un corps à l'autre, c'est à mon sens une telle absurdité qu'elle n'a pu venir à l'esprit d'aucun homme tant soit peu versé dans les choses philosophiques ». L'action à distance impliquait certainement à ses yeux l'influence de «quelque chose», un intermédiaire. En partant du principe d'inertie, fondamental en physique, deux corps isolés n'agissant pas l'un sur l'autre, et sur lesquels n'agit aucune « force », restent en repos relatif ou gardent la même vitesse relative. On ne peut altérer le mouvement d'un corps qu'en faisant interagir un autre corps dont le mouvement sera altéré en retour; l'action égale la réaction. Il en découle que la matière et le mouvement de la matière se conservent. Concrètement, la notion de force ou action est ainsi directement apparentée à l'action d'un corps sur un autre corps, et partant de cette interprétation, il est facile d'imaginer qu'un corps accéléré (accélération gravitationnelle) est en interaction avec ce « quelque chose» de Newton (un autre corps indétectable?). L'accélération gravitationnelle pourrait-elle impliquer l'action d'un autre corps; de corpuscules? Comment concevoir l'attraction alors que le choc d'un corps sur un autre est une poussée et non pas une attraction? L'explication objective d'un mode d'action compréhensible et simple est contenue dans la théorie corpusculaire formulée par Lesage vers le milieu du 18e siècle [1]. On y suppose que « dans les espaces intersidéraux, circulent dans tous les sens, à très grandes vitesses, des corpuscules très ténus. Un corps isolé dans l'espace ne sera pas affecté, dans son mouvement, par les chocs de ces corpuscules, puisque ces chocs se répartissent également dans toutes les directions. Mais, si deux corps A et B sont en présence, le corps A jouera le rôle d'écran et interceptera une partie des corpuscules qui, sans lui, auraient frappé B. Les chocs reçus par B dans la direction opposée à celle de A ne seront plus que partiellement compensés et ils pousseront B vers A » (Fig. 1).

Ce raisonnement n'est pas sans difficultés, mais, il est d'autant plus intéressant qu'il n'implique pas de nouveaux principes et tente d'expliquer le mécanisme qui pourrait se cacher derrière la loi d'attraction. Dans le pur respect des lois d'inertie, le phénomène gravitationnel devient la conséquence de l'interaction entre la matière « palpable » d'un corps et celle « indétectable » du milieu corpusculaire. L'apparente attraction d'un corps sur un autre corps résulterait non pas d'une force fictive d'attraction ou d'une altération géométrique de l'espace, mais de la poussée (l'action non équilibrée) du milieu corpusculaire affecté par la présence des deux corps.

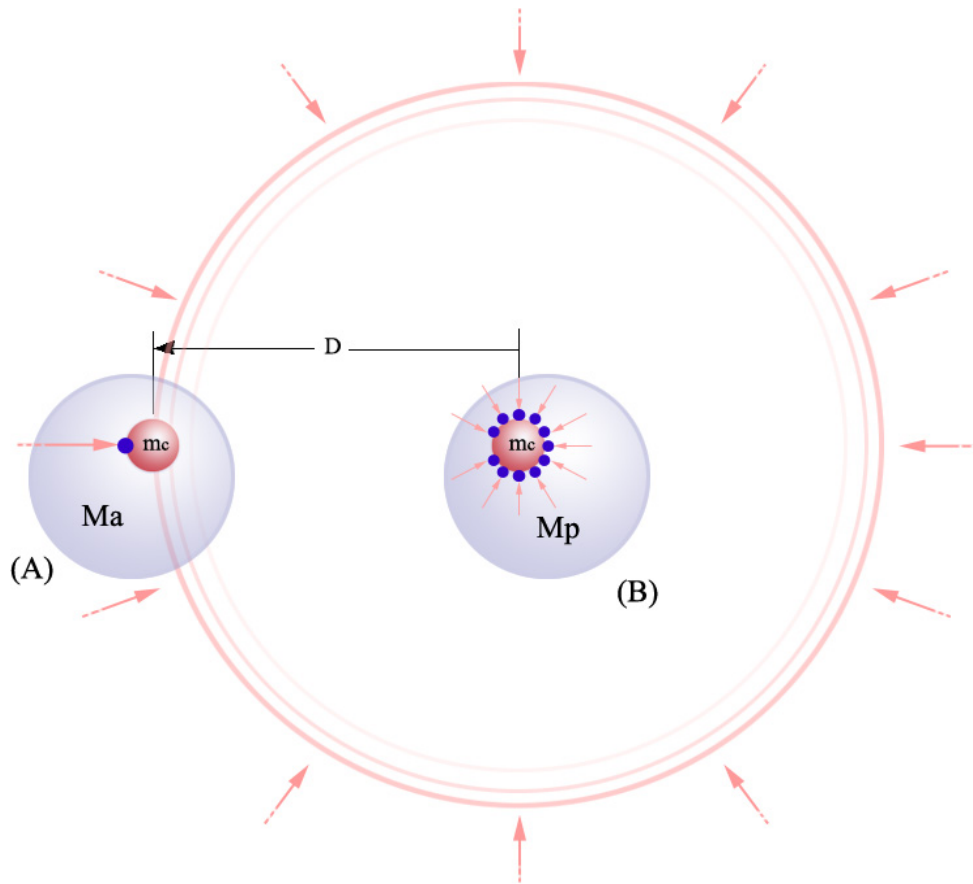

FIG 1. Le corps *B* de masse *Mp* subit une poussée non équilibré du flux corpusculaire en présence du corps *A* de masse *Ma*. On illustre l'interaction entre deux éléments de masse $m_c$ de section efficace d'interaction $\sigma$ et placés à distance *D* l'un de l'autre (en réalité $4\pi D^2$ est très grand devant $\sigma$, et $m_c$ représente la masse « élémentaire » des composantes sub-quantiques dans ce processus d'interaction).

Avec une vision et une connaissance moderne de la structure interne de la matière, le modèle corpusculaire gagne en clarté. Il ne s'agit plus, ici, de l'action de corpuscules à l'échelle des atomes, comme l'interprétait Poincaré [2], mais d'une interaction à un niveau de structure plus intime de la matière, vraisemblablement à l'échelle de pénétration et d'interaction des neutrinos. Ce pouvoir de pénétration de la matière par de fines particules « neutres » n'était pas admis à l'époque de Poincaré, et celui-ci avait fortement critiqué l'hypothèse corpusculaire en s'appuyant principalement sur cet argument d'époque. Pourtant avec les neutrinos, on peut reconnaître aujourd'hui que le degré de pénétration requis par la théorie corpusculaire puisse exister. Ce vieux concept est encore aujourd'hui la seule explication concrète de ce qui est peut-être la vraie nature du champ gravitationnel. Dans cette étude, la théorie corpusculaire a été développée sur une base quantitative afin d'être confrontée aux faits expérimentaux et aux données actuelles.

## 1.1 Calcul de l'attraction gravitationnelle

Nous pouvons jeter les bases d'une description théorique en supposant simplement que les corpuscules gravitationnels, que nous appelons gravitons, transfèrent leur quantité de mouvement à la matière par absorption. On peut démontrer que pour obtenir un effet d'attraction, la théorie corpusculaire impose que les chocs soient au moins en partie inélastiques. Le plus simple est de considérer que l'absorption se produit pour chaque collision avec les éléments constitutifs des particules élémentaires (composantes sub-quantiques) et que ces éléments absorbants ont une masse $m_c$ et une section efficace d'interaction $\sigma$ vis à vis des gravitons.

Le flux d'impulsion (flux équilibré) qui atteint un élément absorbant peut être défini par

$$\vec{Io} = (\sum m_g \vec{v}_g \, / \, dt) \, / \, m_c \tag{1}$$

où $m_g$ et $\vec{v}_g$ sont respectivement la masse et la vitesse des gravitons provenant de toutes les directions et absorbés dans $m_c$ pendant un intervalle de temps $dt$. Dans cette relation où $\vec{Io}$ a les dimensions d'une accélération, nous présumons que la masse des gravitons absorbés par unité de temps est négligeable devant $m_c$ ; soit $\Sigma \, m_g \, / dt \; \ll \; m_c$.

Prenons un corps de masse « active » $Ma$ situé à distance $D$ d'un autre corps de masse « passive » $Mp$. Un élément $m_c$ de $Mp$ absorbe tout le flux centré sur lui, il subit donc un ensemble d'impulsions sous l'effet des impacts des gravitons (Fig.1). En l'absence de $Ma$ l'effet global est nul ( $\vec{Io} = 0$ ) car la somme vectorielle des impulsions est nulle; les impacts étant uniformément distribués tout autour de $m_c$.

En présence d'une masse active $Ma$, le flux global $Io$ susceptible d'atteindre un élément absorbant de $Mp$ va d'abord être affecté par les éléments absorbants de $Ma$. Chaque élément de $Ma$ situé à la distance $D$ de $Mp$ produira donc une diminution $\sigma(Io \, / 4\pi D^2)$ du flux $Io$ dirigé vers un élément $m_c$ de $Mp$. En utilisant ( $Io \, / 4\pi D^2$), nous définissons la densité de flux par unité de surface en tout point de la sphère, de rayon D, centrée sur l'élément « passif ». On conçoit facilement que $\sigma$ (appartenant à l'élément « actif ») n'est qu'un point sur cette sphère. Ainsi, l'accélération subie par $Mp$ (ou par chacun de ses $m_c$ éléments « passifs ») à cause de l'influence sur le milieu corpusculaire des ($Ma \, / mc$) éléments « actifs », suffisamment distants et regroupés pour être supposés concentrés en un point à la distance $D$, sera

$$g \; = (\sigma(\, Io \; / 4\pi D^2 \,)) \; (\, Ma \, / \, m_c \,)$$

En modifiant l'arrangement des termes on obtient

$$g \; = \left[ \frac{Io}{4\pi} \frac{\sigma}{m_c} \right] \frac{Ma}{D^2} \tag{2}$$

L'influence de $Ma$ sur $Mp$ étant un effet soustractif qui agit sur le flux extérieur, l'accélération $g$ de $Mp$ est un vecteur dirigé vers $Ma$. L'expression 2 est équivalente à la loi d'attraction de Newton où la constante de gravitation universelle

$$G \; = \left[ \frac{Io}{4\pi} \frac{\sigma}{m_c} \right] \tag{3}$$

La valeur de $G$ est directement proportionnelle au flux corpusculaire et à la section efficace d'interaction par unité de masse pour les interactions gravitationnelles.

## 1.2 Atténuation gravitationnelle et correction à la loi de Newton

La théorie corpusculaire conduit à une loi d'attraction en inverse de la distance au carré, lorsqu'on assume une indépendance d'action des éléments absorbants par rapport à un flux corpusculaire extérieur uniforme. Cependant, lorsqu'il s'agit d'une très grosse masse comme le Soleil, chaque élément intercepte un flux déjà réduit par la présence des autres éléments. Dans ce cas, l'effet gravitationnel effectif dépend, pour chaque direction, de la densité du flux résiduel qui traverse l'élément actif. L'absorption cumulative qui réduit plus ou moins la densité de flux au niveau d'un élément absorbant, et par conséquent la valeur effective et directionnelle de $G$, va introduire une correction à la loi d'attraction gravitationnelle donnée par la relation 2.

Si nous prenons un élément de masse $dM$ compris dans l'angle solide $d\beta$ et situé à distance $L$ de $Mp$, alors $dM = \rho\,(\pi/4)\,L^2\,d\beta^2\,dL$, où $\rho$ est la densité de l'élément $dM$ (voir Fig. 2, où $Mp$ est identifié à une particule d'épreuve ou au centre de masse d'une planète). Le nombre d'éléments absorbants dans $dM$ est $dM/m_c$. On calcule $f$, l'atténuation du flux causée par $dM$, en prenant la somme ($A$) des sections efficaces de ses éléments absorbants divisée par la surface totale ($S$) du flux incident.

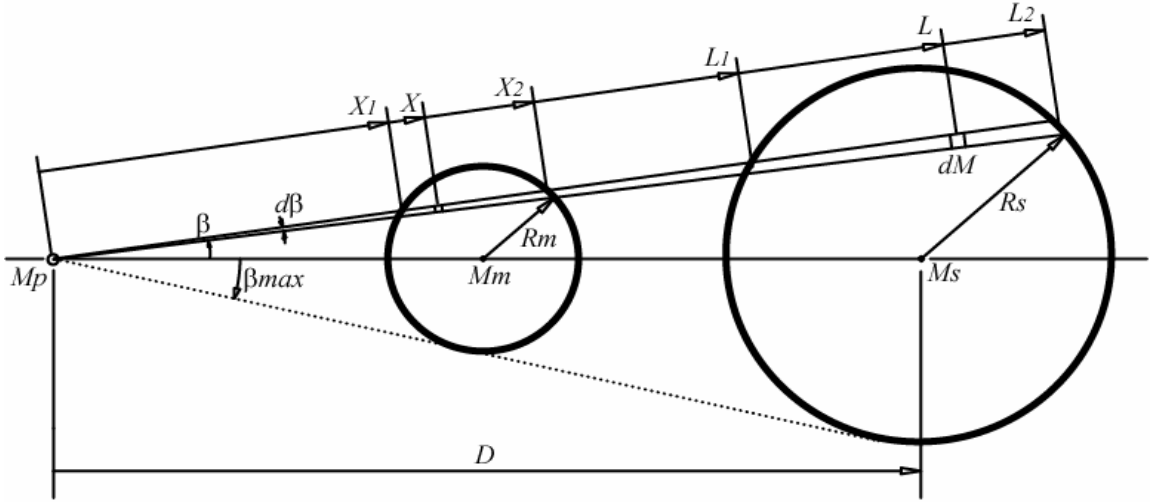

FIG 2. Repérage géométrique des éléments de masse de la Lune et du Soleil relativement à une particule d'épreuve à $Mp$. Les points $Mm$ et $Ms$ correspondent respectivement au centre de masse de la Lune et du Soleil.

Avec $A = (dM/m_c)\sigma$ et $S = (\pi/4)\,L^2\,d\beta^2$, nous obtenons après réduction

$$f = K\,\rho\,dL \qquad\qquad (4)$$

$$K = \frac{\sigma}{m_c} \qquad\qquad (5)$$

Pour simplifier l'écriture, nous avons introduit la « constante K » qui est la section efficace d'interaction par unité de masse pour les gravitons.

Si pour l'élément de masse $dM$, nous connaissons la fraction $P$ du flux Io préalablement absorbée par les autres éléments de masse situés entre $L_2$ et $L$, l'atténuation effective relativement au flux Io, dans la direction considérée, sera

$$fr = f(1-P) \qquad (6)$$

$P$ est donnée par la somme des $fr$ provenant des éléments de masse situés entre $L_2$ et $L$, soit

$$P = \int_0^{L_2-L} fr \qquad (7)$$

En combinant (4), (6) et (7) on obtient

$$P = \int_0^{L_2-L} K\rho(1-P)\,dL \qquad (8)$$

qui a pour solution, si on suppose une <u>densité uniforme</u>,

$$(1-P) = e^{-K\rho(L_2-L)} \qquad (9)$$

Cette dernière relation donne la correction à apporter pour obtenir le flux résiduel $I$, qui sera présent au niveau de l'élément de masse $dM$. La correction dépend de la densité $\rho$ de la matière traversée par le flux extérieur et de l'épaisseur ($L_2$ - $L$) pour atteindre $dM$ ; elle s'applique relativement au flux dirigé vers la masse passive (ou particule d'épreuve). L'influence gravitationnelle de $dM$, qui est proportionnelle au flux qu'il affecte, sera ainsi réduite. La relation 9 caractérise l'atténuation du flux corpusculaire Io à travers la matière. Avec une masse importante comme le Soleil, on doit en tenir compte dans les calculs d'attraction gravitationnelle. Aussi, pour obtenir l'attraction du Soleil sur une particule d'épreuve, on doit calculer la somme des effets gravitationnels des éléments de masse en multipliant $G$ par $e^{-K\rho(L_2-L)}$ dont la valeur dépend, pour chaque élément de masse, de la distribution de matière solaire par rapport à $Mp$. Le terme correcteur utilisé dans cette étude permet de calculer correctement l'effet d'atténuation gravitationnelle, mais on doit normalement tenir compte de la variation de densité entre $L_2$ et $L$ (ex : variation radiale de densité avec les modèles solaires proposés).

## 2 – Correction appliquée à l'avance du périhélie des planètes

En évaluant, à la figure 2, la valeur que peut prendre ($L_2$ - $L$) pour un élément de masse $dM$ quelconque, on constate que cette épaisseur traversée jusqu'à $dM$ devient plus grande lorsque $Mp$ s'éloigne de $Ms$. Par conséquent, la valeur prise par $G\,e^{-K\rho(L_2-L)}$ va diminuer lorsqu'une particule d'épreuve passe du périhélie à l'aphélie. L'atténuation gravitationnelle va donc produire, pour chacun des éléments de masse du Soleil, une diminution de l'attraction avec la distance légèrement plus grande que celle prévue par la loi d'attraction en $1/r^2$. Cette conséquence introduit une correction qui s'accorde qualitativement avec une avance du périhélie des planètes. Un test décisif avant de vouloir réhabiliter la théorie corpusculaire est d'abord de vérifier si elle peut tenir compte quantitativement de l'avance du périhélie des planètes.

## 2.1 Calcul de la section efficace d'interaction

Ce calcul de la section efficace d'interaction $K$ s'appui sur l'écart observé de l'avance du périhélie de Mercure par rapport à l'avance prévu dans les calculs de mécanique céleste. Pour en faciliter l'évaluation et afin de comparer nos résultats avec la théorie d'Einstein de la gravitation, nous avons procédé par simple analogie. Les corrections apportées à la loi de Newton par la théorie corpusculaire ou par la théorie d'Einstein peuvent être absorbées dans G, aussi elles ont été comparées sur la base d'un effet équivalent à une variation linéaire de la "constante" gravitationnelle $G$ entre le périhélie et l'aphélie de Mercure. Si les variations produites sur G pour les deux théories sont de même grandeur et de même signe, il est logique d'assumer que, leur effets sur l'avance du périhélie seront les mêmes. On peut montrer que l'approximation linéaire est acceptable entre ces deux positions orbitales.

Dans le développement qui suit, nous utilisons la loi de Newton corrigée pour l'atténuation. Afin de simplifier les calculs numériques (voir annexe 1), dans cette première évaluation nous avons considéré une densité uniforme pour le Soleil. Prenons $dM$, une portion annulaire de matière solaire repérée à la Fig. 2 par ses coordonnés $L$ et $\beta$, cette matière solaire se trouve à la même distance $L$ de $Mp$, produit une composante effective d'attraction proportionnelle à $cos\beta$ et est affectée par la même atténuation $e^{-K\rho(L_2-L)}$. Pour calculer l'attraction du Soleil, il faut évaluer l'épaisseur ($L_2 - L$) de matière traversée jusqu'à $dM$ dans la direction du flux de gravitons passant par $dM$ et $Mp$. L'accélération produite à $Mp$ par le Soleil est obtenue en additionnant les actions gravitationnelles de l'ensemble des éléments $dM$. La loi d'attraction corrigée s'écrit

$$g = \int_0^{\beta_{max}} \int_{L_1}^{L_2} G e^{-K\rho(L_2-L)} \rho \, 2\pi \, L^2 \sin(\beta) \cos(\beta) \frac{1}{L^2} \, dL \, d\beta \qquad (10)$$

où $\rho$ est la densité moyenne du Soleil et $L$ est la distance entre $dM$ et $Mp$. Les valeurs $L_1$, $L_2$ et $\beta_{max}$ sont des fonctions de $Rs$, $\beta$ et $D$ ($Rs$ est le rayon du Soleil et $D$ est la distance entre $Mp$ et le centre du Soleil). L'accélération sans atténuation (loi de Newton) est

$$g_n = \int_0^{\beta_{max}} \int_{L_1}^{L_2} G \, \rho \, 2\pi \, L^2 \sin(\beta) \cos(\beta) \frac{1}{L^2} \, dL \, d\beta \qquad (11)$$

qui est équivalent à $g_n = G \dfrac{M_s}{D^2}$ pour une masse sphérique comme le Soleil.

La variation $\Delta G$ de la "constante" gravitationnelle $G$ entre le périhélie et l'aphélie due au seul effet d'atténuation est évalué par

$$\Delta G = \frac{\left[\left(g_p\right) D_p{}^2 - \left(g_{pa}\right) D_p{}^2\right]}{M_s} \qquad (12)$$

où $g_p$ a été calculée, avec la relation 10, pour la distance $Dp$ correspondant au périhélie de Mercure; et $g_{pa}$ a été obtenue en remplaçant les valeurs d'atténuation du périhélie par celles que nous aurions à l'aphélie. Contrairement à l'expression 11, la présence des effets locaux d'atténuation nous empêche de simplifier l'expression 10, et d'attribuer la variation de $g$ comme étant due à la seule variable $D$. On arrive de cette manière à estimer la variation moyenne de $G$ en procédant à partir d'une même distance, celle du périhélie. D'une manière générale on calcule $\Delta G$ en faisant la différence pour une même distance orbitale entre les constantes $G$ moyennes calculées avec l'effet de l'une ou l'autre atténuation (périhélie et aphélie). On assume que cette manière d'estimer $\Delta G$ est acceptable; d'ailleurs on ne trouve qu'une légère différence, inférieure à 0,3%, entre les valeurs calculées, selon que l'on procède à partir du périhélie ou de l'aphélie (voir annexe 1).

Le terme correcteur d'Einstein $\dfrac{3 G M_s}{c^2 D^2}$ pour étudier la trajectoire d'une particule d'épreuve dans un champ de Schwarzschild [3] est également évalué par comparaison à une variation $\Delta G$ de la "constante" gravitationnelle qu'il introduirait entre le périhélie et l'aphélie d'une planète. Ainsi, en absorbant ce terme dans G nous obtenons pour la théorie d'Einstein,

$$\Delta G = \frac{12 G^2 M_s}{c^2} \frac{e}{a(1 - e^2)} \tag{13}$$

où $c$ est la vitesse de la lumière, $M_s$ est la masse du Soleil, $e$ est l'excentricité de l'orbite et $a$ est le demi grand-axe de l'orbite.

Lorsqu'on fait l'égalité entre (12) et (13) pour Mercure, on obtient par calcul numérique une section efficace d'interaction

$$K = 3.2 \times 10^{-17} \, \text{m}^2/\text{kg} \tag{14}$$

En fait, pour évaluer $K$ d'une manière plus conventionnelle, nous pourrions aussi tenter d'extraire le terme correcteur dû à l'atténuation dans l'expression (10) et lui appliquer une démonstration similaire à celle utilisée pour le terme correcteur d'Einstein, dans la référence 7, en sorte de retrouver l'avance du périhélie de Mercure. La méthode qui a été adoptée dans la présente étude est équivalente et en pratique beaucoup plus simple.

Notre valeur de $K$, à priori strictement proportionnelle à la masse, correspond à une section efficace de $5 \cdot 10^{-40}$ cm$^2$ /nucléon et de $3 \cdot 10^{-43}$ cm$^2$ /électron. Il s'agit d'un résultat assez remarquable, car bien qu'il n'ait pas de lien direct avec la mécanique quantique, il tombe précisément dans le domaine d'interaction des neutrinos avec la matière. On rapporte couramment dans la littérature des sections efficaces d'interaction dans ces ordres de grandeur pour des neutrinos de quelques dizaines de MeV.

## 2.2 Calcul de l'avance du périhélie des planètes

La valeur de $K$ étant déterminée (14), on peut calculer l'avance du périhélie des autres planètes en utilisant les expressions (10) et (12). On obtient l'avance du périhélie, en radian, en multipliant par $(\pi / (2 G e))$ la valeur de $\Delta G$ calculée avec (12) ou (13).
Le tableau I donne l'avance du périhélie des planètes en secondes d'arc par siècle (voir les calculs à l'annexe 1). Les valeurs calculées avec la théorie corpusculaire sont en très bon accord avec les valeurs observées pour l'avance du périhélie des planètes. Les termes correcteurs sont très différents dans les deux théories, l'un est additif et s'applique à la masse centrale, tandis que l'autre est soustractif et s'applique à chaque élément de masse du Soleil. Il est pourtant curieux de constater que les deux théories nous donnent sensiblement les mêmes résultats pour l'ensemble des planètes.
Nous avons repris les calculs précédents en additionnant les effets avec un modèle à deux couches de densité homogènes; soit un noyau de 0.25Rs renfermant 40% de la masse solaire avec une enveloppe périphérique. Cette simplification, qui nous rapproche du modèle solaire standard, nous redonne les valeurs du tableau 1, mais avec une valeur $K$ environ deux fois plus faible. On peut raisonnablement croire qu'un calcul précis utilisant la variation de densité radiale du modèle solaire va nous donner une section efficace d'interaction d'environ $1 \cdot 10^{-17}\,m^2/kg$.

TABLEAU I. Avance du périhélie des planètes

| Planète | $\Delta\Omega$/siècle | | |
|---------|-------------|---------------|-----------------|
|         | Einstein [3] | Corpusculaire | Expérimental [3] |
| Mercure | 42.95" | 43.07" | 43.11"±0.45" |
| Vénus | 8.61" | 8.63" | 8.4"±4.8" |
| Terre | 3.83" | 3.84" | 5.0"±1.2" |
| Mars | 1.35" | 1.35" | Non disponible |

## 2.3 Comparaison entre la loi corrigée et la loi de Newton

Le fait d'introduire l'atténuation gravitationnelle, dans l'expression 10, nous oblige à apporter une petite correction d'environ 0.0024% à la masse estimée du Soleil, afin de retrouver la valeur Newtonienne de l'attraction du Soleil à une Unité Astronomique. Le terme supplémentaire, dans l'expression 10, introduit en plus de la courbure, un certain décalage par rapport à la loi de Newton (expression 11); aussi lorsqu'on fait coïncider les deux fonctions à 1 U.A., la loi corrigée donne une valeur d'attraction légèrement supérieure aux plus grandes distances. L'écart avec la loi d'attraction de Newton tend rapidement vers sa valeur asymptotique et reste inférieur à une partie par milliard pour l'ensemble des distances planétaires. La théorie corpusculaire permet donc de prendre en compte l'avance du périhélie des planètes sans introduire de désaccord avec les calculs de mécanique céleste.

# 3 – Anomalie gravitationnelle par occultation

Au cours d'une éclipse solaire, quand la Lune passe devant le Soleil, elle se retrouve dans un cône d'influence où le flux corpusculaire dirigé vers la zone de l'éclipse a été réduit par le Soleil. Puisque l'effet d'attraction de la Lune est proportionnel au flux qu'elle affecte, en interceptant un faisceau corpusculaire déjà réduit, la lune aura une influence attractive réduite pour tout observateur situé dans la zone de l'éclipse. Bien que le phénomène soit simple, un calcul précis de cet effet d'occultation, variable dans le temps, est assez complexe car le phénomène est combiné à la marée terrestre qui est la résultante de l'attraction locale et de l'accélération d'entraînement global de la Terre.

Cet effet direct de l'occultation représente donc une diminution de l'attraction lunaire pendant une éclipse de Soleil. Il en résulte une augmentation locale de la pesanteur qui se traduit par une anomalie gravitationnelle positive.

L'effet maximal d'occultation sur l'attraction lunaire est évalué en prenant des densités moyennes, $\rho_m$ pour la Lune et $\rho$ pour le Soleil, et en assumant un recouvrement exact de leur disque apparent pendant l'éclipse. Cette condition de recouvrement avec celle de densité homogène facilite grandement les calculs pour nous permettre d'évaluer l'ordre de grandeur du phénomène. L'attraction lunaire au moment culminant de l'éclipse est

$$g_{me} = \int_0^{\beta_{max}} \int_{X_1}^{X_2} G \left[ e^{-K \, \rho_m \, (X_2 - X)} \right] \left[ e^{-K \, \rho \, (L_2 - L_1)} \right] \rho_m \, 2\pi \, X^2 \, \cdots$$
$$\cdots \, \sin\beta \cos\beta \, \frac{1}{X^2} \, dX \, d\beta \qquad (15)$$

où $X$, $X_1$, $X_2$, $L_1$, $L_2$ et $\beta_{max}$ sont explicités à la figure 2. L'attraction sans l'atténuation du Soleil est

$$g_m = \int_0^{\beta_{max}} \int_{X_1}^{X_2} G \left[ e^{-K \, \rho_m \, (X_2 - X)} \right] \rho_m \, 2\pi \, X^2 \sin\beta \cos\beta \, \frac{1}{X^2} \, dX \, d\beta \qquad (16)$$

L'attraction lunaire doit donc subir une diminution relative égale à $(g_m - g_{me}) / g_m$. Un calcul numérique avec (14), (15) et (16) donne une diminution locale de 0,005% de l'attraction lunaire, et par conséquent de sa composante verticale d'attraction, pendant une éclipse totale de Soleil. Cet effet local traverse entièrement la Terre dans la prolongation du cône d'ombre, mais décroît latéralement vers la zone de pénombre. Il est facile d'estimer qu'il n'affecte qu'une faible fraction de la masse terrestre.

## 4 – Résultat de l'éclipse de Soleil du 10 mai 1994

Lorsque la Lune et le Soleil sont au dessus de l'horizon, leur influence gravitationnelle est négative par rapport à l'attraction terrestre. En effet, à ce moment, l'action verticale qu'exercent ces astres pour un observateur terrestre est dirigée vers le haut et agit en sens opposé à l'attraction terrestre. Au moment culminant de l'éclipse de Soleil du 10 mai 1994 [4], à 13 h 38, près de Montréal, latitude 45°30'.00N, longitude 73°30'.00W, la composante verticale de l'accélération gravitationnelle due à la Lune était estimée à -2.66 mgal. Cette valeur est donnée par $G \, \frac{M_m}{X^2} \sin\alpha$, où $M_m$ est la masse lunaire, $X$ est la distance entre la Lune et le lieu de la mesure et $\alpha$ est l'altitude de la Lune. L'altitude lunaire varie de 62.2° à 49.2° durant l'éclipse [5] alors que la distance $X$ change très peu, par conséquent la composante verticale de l'accélération lunaire reste comprise entre -2.7 et -2.0 mgal.

La variation diurne de la pesanteur doit normalement être considérée dans les relevés gravimétriques, aussi, la Commission Géologique du Canada (CGC) fournit des tables de correction de marée pour la prospection géophysique. Les mesures du 10 mai 1994, effectuées par une firme de géophysique spécialisée dans les relevés micro-gravimétriques, ont été obtenues à l'aide d'un gravimètre LaCoste & Romberg, modèle D, installé sur une station fixe [6]. Dans ces conditions idéales, le type de gravimètre utilisé permet de prendre des mesures relatives avec une erreur de fidélité aussi faible que ± 0.5 μgal.

La figure 3 donne la variation de l'accélération de la pesanteur (valeurs relatives en mgal ) en fonction de l'heure de la mesure. Une fonction sinusoïdale permet l'ajustement des corrections de marée tabulées (entre 11 h 00 et 16 h 00, heure locale) par la CGC pour Montréal. La courbe en trait interrompu, dont le niveau a été ajusté sur le premier point, correspond à ces corrections. La fonction de lissage suivante

$$115.4835 + 0.1135(\sin(0.429 H_r - 0.75))$$

où $H_r$ est l'heure locale, a donc été appliquée aux points expérimentaux obtenus en dehors de la période de l'éclipse. Cette fonction de régression de la valeur relative de l'accélération de la pesanteur en fonction de l'heure locale devait permettre de faire ressortir correctement toute discontinuité, qui pourrait résulter de l'occultation du Soleil par la Lune, et absorber la dérive instrumentale, qui présente une variation continue dans le temps. La dérive instrumentale est appréciée en comparant la courbe de régression avec celle tirée des corrections de la CGC. La courbe de régression, en trait continu, a été superposée aux points expérimentaux. Les droites verticales indiquent le début et la fin de l'éclipse visuelle. Les valeurs expérimentales montrent une anomalie bel et bien repérable et en parfaite coïncidence avec la période de l'éclipse.

La figure 4 donne la différence en µgal entre les valeurs expérimentales et celles calculées par la fonction de régression. L'erreur-type d'estimation est inférieure à 0.6µgal pour les points ajustés. L'ensemble des points correspondant à la période de l'éclipse présente un décalage positif d'environ 4 erreur-types, aussi d'un point de vue statistique une anomalie est présente. Puisqu'elle n'a pu être reliée à une erreur instrumentale dû à une variation de pression ou de température ambiante, on peut raisonnablement croire qu'elle est attribuable à un effet gravitationnel résultant de l'occultation du Soleil par la Lune. En effet, le gravimètre est maintenu à 51°C dans une enceinte isolée et régulée en température à <0.3°C, de sorte que l'influence de la variation de température ambiante est négligeable. L'enceinte est également scellée pour l'isoler des effets de variation de pression (poussée d'Archimède). On rapporte pour ce type d'appareil une influence de la pression atmosphérique de moins de 0.4 µgal/milibar [8].

Le tableau 2 donne les variations de pression atmosphérique, dans la région de Montréal, fournies par Environnement Canada. On peut en conclure que la pression atmosphérique ne peut être responsable de l'anomalie, puisque la variation de pression maximale (+0.7 milibar) ne coïncide pas avec le phénomène et ne peut tenir compte que du 1/10 de sa valeur.

TABLEAU 2. Variation de la pression atmosphérique à Montréal le 10 mai 1994

| Heure locale | 11 | 12 | 13 | 14 | 15 | 16 | 17 |
|---|---|---|---|---|---|---|---|
| Pression (milibar) | 1012,7 | 1012,5 | 1012,5 | 1013,2 | 1013,2 | 1013,2 | 1013,1 |

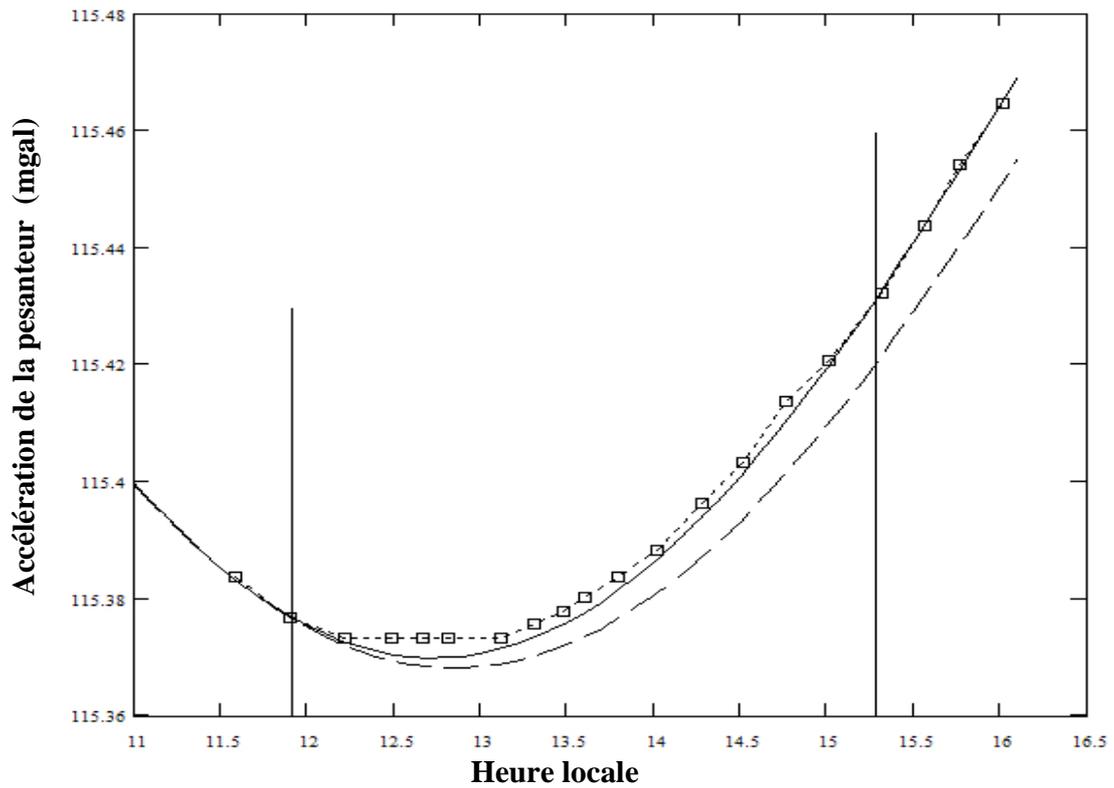

FIG 3. Valeurs relatives de l'accélération de la pesanteur (mgal) en fonction de l'heure locale, mesurées le 10 mai 1994 à Boucherville, Québec, Canada. Les lignes verticales permettent de repérer dans le temps le début et la fin de l'éclipse de Soleil. La courbe de régression en trait continu (sinusoïde des moindres carrés pour les valeurs en dehors de la période de l'éclipse) donne l'accélération de la pesanteur sans l'anomalie [7]. La courbe en trait interrompu correspond à la marée terrestre déduite des tables de corrections de la Commission Géologique du Canada.

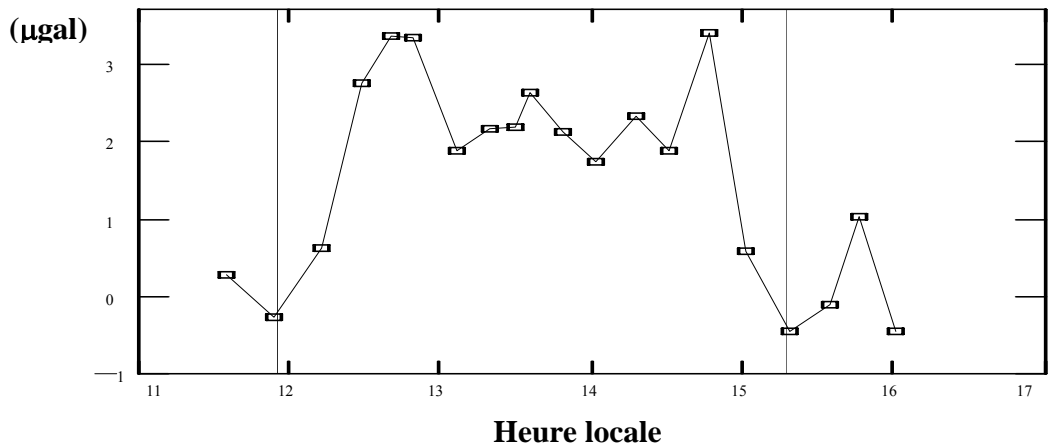

FIG 4. Anomalie gravitationnelle observée pendant l'éclipse de Soleil du 10 mai, 1994 à Boucherville, Québec, Canada. Les points représentent la différence en μgal entre les valeurs expérimentales et les valeurs données par la courbe de régression. La période de l'éclipse visuelle est délimitée par les lignes verticales.

L'anomalie moyenne mesurée pendant l'éclipse est de +2.4 ± 0.5 µgal. Sa réelle association à un phénomène gravitationnel demande à être validé par d'autres mesures en utilisant une instrumentation plus précises, cependant son lien avec l'éclipse est supporté par le fait que l'anomalie coïncide parfaitement avec la période de l'éclipse. L'ordre de grandeur de l'anomalie avait été initialement estimé à une valeur au dessus de la limite de sensibilité instrumentale, c'est pourquoi cette vérification du 10 mai, 1994, a été effectuée [7]. Cependant à cause d'un financement très limité, et à moins d'un mois de l'événement, seul un relevé gravimétrique classique a pu être réalisé. Les conditions pour mettre le phénomène en évidence étaient favorables puisque la Lune était à son altitude maximale pendant l'éclipse. De plus, le caractère annulaire de l'éclipse ne pouvait que maximiser l'effet d'occultation puisque du point de vue de la théorie corpusculaire la Lune se retrouvait presque en totalité dans le cône d'influence maximal du Soleil, à la latitude de Montréal. La composante verticale de l'action lunaire pendant l'éclipse étant de 2.66 mgal, on évalue théoriquement à 0,13µgal (soit 0.005% x 2.66 mgal) l'anomalie causée par l'éclipse de Soleil. Nos mesures gravimétriques du 10 mai 1994 montrent la présence d'une anomalie gravitationnelle de 20 fois supérieure à la valeur théorique.

## 5 – Discussion

Il est intéressant de rappeler les mesures gravimétriques réalisées pendant les éclipses solaires du 24 octobre 1995 [9] et du 9 mars 1997 [10]. Dans ces deux cas, on a utilisé un gravimètre Lacoste-Romberg et observé une anomalie en coïncidence avec la période de l'éclipse. L'ordre de grandeur des anomalies est comparable, mais contrairement à l'anomalie de 1994 ces anomalies, de profils différents, sont négatives. La différence importante entre ces éclipses est l'élévation de la Lune sur le lieu d'observation des anomalies, soit de quelques degrés en 1995, environ 21° en 1997, alors que nous avions 62° en 1994. Si on admet que ces anomalies ne sont pas des phénomènes fortuits, l'interprétation la plus rationnelle est de les associer à un effet gravitationnel local dont la cause reste bien sûr matière à discussion. Sur la base de notre évaluation théorique précédente, aucun des gravimètres utilisés n'avait la précision pour mettre en évidence un effet direct d'occultation, de l'ordre du dixième de microgal, que prévoit la théorie corpusculaire. Parmi les causes conventionnelles pour expliquer une anomalie négative, on a invoqué un effet de la gravité atmosphérique dû à une augmentation de la densité atmosphérique dans le cône d'ombre au dessus du lieu des mesures [11], mais les mouvements de masses d'air que cela implique ne paraissent pas réalistes, ni observables au sol. L'effet gravitationnel le plus vraisemblable reste donc associé à un léger mouvement de la croûte terrestre (soit une inclinaison du sol, un changement de niveau ou une combinaison des deux) qui se serait produit pendant l'éclipse. Les causes conventionnelles provenant de l'éclipse et pouvant produire un tel effet sur la croûte sont reliées à un changement de pression et de température. Même en assumant que la pression atmosphérique puisse avoir un effet significatif, il n'y a pas de corrélation entre les variations de pression atmosphérique et les anomalies de 1994 et 1997 pour lesquelles la pression a été relevée. La variation de température pendant une éclipse a été invoquée, mais reste assez faible et superficielle [10, 11]. Il semble d'ailleurs difficile de concevoir que la variation de la température de surface, affectant généralement les sols meubles, puisse générer des contraintes suffisantes pour induire des dilatations et des distorsions significatives sur la croûte terrestre.

Il est facile d'évaluer l'effet que peut produire un léger abaissement $h$ de la surface de la croûte terrestre sur laquelle repose le gravimètre. La variation de la pesanteur, qui correspond au changement de distance par rapport au centre de la Terre, est

$$G \, M_t \left[ \frac{1}{(R_t - h)^2} - \frac{1}{R_t^{\;2}} \right]$$

où Mt est la masse terrestre et Rt est le rayon terrestre. Si on utilise cette dernière formule avec un affaissement local $h$ de seulement 7 mm on trouve une augmentation de gravité de 2,1 µgal qui est de l'ordre de grandeur de l'anomalie de 1994. On peut se demander si l'accélération théorique de 0,13 µgal prévue par la théorie corpusculaire est capable de produire un affaissement local suffisant pour expliquer l'anomalie. Pour répondre à cette question, nous avons évalué le déplacement parcouru par un corps en chute libre dans un champ de gravité de 0,065 µgal pendant une durée de 1 heure 43 minutes, qui correspond au temps écoulé entre le premier contact et la phase maximale de l'éclipse. Nous avons utilisé la moitié de l'accélération théorique pour tenir compte du fait que l'accélération va croître de zéro jusqu'à sa valeur maximale. Il est logique de croire que si une portion importante de la planète (prolongement intra planétaire du cône d'ombre) est soumise à une accélération différentielle par rapport à l'autre portion, elle puisse être assimilée à un corps en chute libre dans ce différentiel d'accélération; surtout si on considère des déplacements progressifs de l'ordre du millimètre et étalés sur une zone de transition de plusieurs centaines de kilomètres (zone de pénombre). Avec la formule classique $h = \frac{1}{2} a \, t^2$ on trouve un déplacement vers le centre de la Terre de 12,5 mm; donc presque deux fois supérieur à l'affaissement requis pour produire l'anomalie observée en 1994. On peut en conclure que même si le différentiel d'accélération prévu par la théorie corpusculaire est trop faible pour avoir été détecté, il peut induire des déplacements susceptibles de générer des effets mesurables. Il ne s'agit pas uniquement d'un déplacement de la surface sur laquelle repose le gravimètre, puisque le cône d'influence traverse la Terre entière, c'est donc toute la masse interne située dans le prolongement du cône d'ombre et de pénombre qui est affectée. Ce sont les minuscules déplacements relatifs de blocs entiers de la planète qui seraient susceptibles de produire les variations anormales de la gravité au passage de l'éclipse.

En réalité, le phénomène est beaucoup plus complexe que le simple calcul précédent. L'effet réellement induit va dépendre des déplacements de masses et des distorsions qui affecte la position et l'inclinaison du gravimètre par rapport à la distribution de masse terrestre. Cet effet est fonction, entre autres, de la position du cône d'ombre par rapport à l'axe Terre-Lune, du lieu du relevé gravimétrique, de l'heure et la durée de l'éclipse. La conception de l'appareil utilisé et son orientation spatiale sont également d'autres facteurs qui peuvent influencer le résultat. Quoiqu'il en soit, l'atténuation gravitationnelle d'éclipse se présente comme une excellente candidate pour expliquer quantitativement l'anomalie d'éclipse de 1994.

La dynamique des déplacements combinés à la rotation terrestre et à la rigidité de la croûte pourrait, en partie, être la cause de la disparité dans les observations d'une éclipse à l'autre; anomalie négative ou positive, forme des anomalies et absence d'anomalie. Modéliser un tel phénomène, combiné à la marée terrestre, afin d'évaluer l'anomalie correspondant à un cas particulier est une tâche assez complexe. En partant des informations disponibles nous serions cependant en mesure de faire des études de cas.

Avec des relevés gravimétriques suffisamment précis, réalisés dans des conditions identiques dans l'axe du cône d'ombre, et des deux côtés de la planète, on pourrait éprouver cette explication non conventionnelle, comme cause responsable des anomalies gravitationnelles d'éclipse. En effet, les déplacements induits, prévus par la théorie corpusculaire, traversent la planète dans le prolongement du cône d'ombre et devraient se retrouver également de l'autre côté de la planète. Ce lieu, côté nuit, serait un endroit privilégié pour réaliser des mesures gravimétriques, loin de l'éclipse visuelle, sans aucunes perturbations produites par le rayonnement solaire.

# 6 – Conclusion

La théorie corpusculaire proposée par Lesage vers le milieu du 18$^e$ siècle a subi une évaluation sommaire par les scientifiques du début du 20$^e$ siècle (H. Poincaré, 1924) avant d'être abandonnée. Le faible niveau de développement qu'on lui avait donné et les connaissances physiques de l'époque favorisaient grandement ses détracteurs. Condamnée sur des présomptions plutôt que sur des faits, la théorie corpusculaire reste valable et ne peut être sérieusement écartée que si ses prédictions sont irréconciliables avec les faits expérimentaux. Les nombreuses observations d'anomalie gravitationnelle d'éclipse, dont celle de mai 1994 appuyée par ce développement théorique, vont dans le sens d'une accréditation de la théorie corpusculaire.
La théorie corpusculaire, théorie explicative par excellence, tente de décrire la nature du champ gravitationnel en ne faisant appel qu'aux lois d'inertie. En effet, la conservation de la quantité de mouvement dans l'interaction de la matière avec le milieu corpusculaire explique de manière naturelle le mécanisme qui conduit à une loi d'attraction en inverse de la distance au carré. Pour la théorie corpusculaire, le champ gravitationnel n'est pas généré (condition sine qua non d'une croissance à l'infini) mais altéré (atténué) par la matière. L'intensité du champ gravitationnel atteint une valeur maximale déterminée par le flux corpusculaire qui ne peut que décroître à travers la matière.

On comprend facilement l'égalité entre masse gravifique et masse inertielle si on suppose que la section efficace d'interaction de nos gravitons avec la matière est strictement proportionnelle à la masse. Une telle proportionnalité impose cependant que l'échelle de grandeur à laquelle se situent les interactions soit d'un niveau sub-quantique. La section efficace d'interaction gravitationnelle qui a été calculée est d'ailleurs comparable à celle des interactions des neutrinos avec la matière. Cette relation qui rapproche le phénomène gravitationnel de nos connaissances en microphysique n'est probablement pas qu'une simple coïncidence. En considérant l'effet d'atténuation et une section efficace d'interaction de $10^{-17}$ m$^2$/kg, on peut démontrer que pour atteindre la limite de validité du principe d'équivalence à l'échelle du laboratoire, telle que tentée dans les expériences d'Eötvös (masses de diverses densités), il faudrait avoir une précision relative meilleure que $10^{-15}$. Une telle précision, très difficilement accessible, n'a pas encore été atteinte dans ce type d'expérience [12].
Le flux d'impulsion corpusculaire Io est atténué de manière classique en traversant la matière et le calcul de l'atténuation nous précise la correction qui doit être apportée à la loi de Newton. L'effet d'atténuation, qui devient sensible pour une grosse masse comme le Soleil, fournit une explication simple et conduit à un accord quantitatif pour l'avance du périhélie des planètes.

Les déformations du géoïde, causées par un effet d'atténuation de la gravité soli-lunaire pendant une éclipse de Soleil, semblent être capable d'expliquer la diversité des observations pour les anomalies gravitationnelles d'éclipse qui se superposent à la marée terrestre.

Le présent développement fournit une base théorique suffisamment précise et consistante pour servir de piste à un certain nombre d'évaluations théoriques et expérimentales. Il conduit à de telles retombées en physique, qu'on ne saurait trop insister pour que la communauté scientifique veuille entreprendre des tests de validation, non seulement sur les anomalies gravitationnelles d'éclipse, mais aussi sur le principe d'équivalence en tenant compte de l'effet d'atténuation de la gravité.

## Références

**DETERMINATION OF THE INTERACTION CROSS SECTION  K
AND THE ADVANCE OF THE PERIHELION OF PLANETS**

## a) Data (MKS)

**Mass of the Sun**  $Ms := 1.989 \cdot 10^{30}$  kg

**Gravitational constant**  $G := 6.668 \cdot 10^{-11}$  $m^3 \cdot kg^{-1} \cdot s^{-2}$

**Radius of the Sun**  $Rs := 6.96 \cdot 10^8$  m

**Speed of light**  $c := 2.998 \cdot 10^8$  $m \cdot s^{-1}$

**Mean density of the Sun**  $\rho := \dfrac{3 \cdot Ms}{4 \cdot \pi \cdot Rs^3}$  $\rho = 1408$  $kg \cdot m^{-3}$

| Orbital parameters of the planets | Semi-major axis  (Dga) _in meter_ | Excentricity  (e) | Orbital period  (T) _in sidereal days_ |
|---|---|---|---|
| Mercury | $Dsm := 57.91 \cdot 10^9$ | $esm := 0.20563$ | $Tm := 87.97$ |
| Venus | $Dsv := 1.0826 \cdot 10^{11}$ | $esv := .0068$ | $Tv := 224.7$ |
| Earth | $Dst := 1.4968 \cdot 10^{11}$ | $est := 0.01673$ | $Tt := 365.25$ |
| Mars | $Dsms := 2.2806 \cdot 10^{11}$ | $esms := .0933$ | $Tms := 686.98$ |

## b) Correction to the law of Newton within the framework of the corpuscular  theory

We have to register the provided values  Dga, e, T  ;  **For studied planet**

Semi-major axis (m)   Exentricity   Period (s.d.)

$Dga := Dsm$   $e := esm$   $T := Tm$

**_Planet distance to the Sun at the perihelion (m)_**   $Dp := Dga \cdot (1 - e)$   $Dp = 4.6002 \cdot 10^{10}$

**_Planet distance to the Sun at the aphelion (m)_**   $Da := Dga \cdot (1 + e)$   $Da = 6.9818 \cdot 10^{10}$

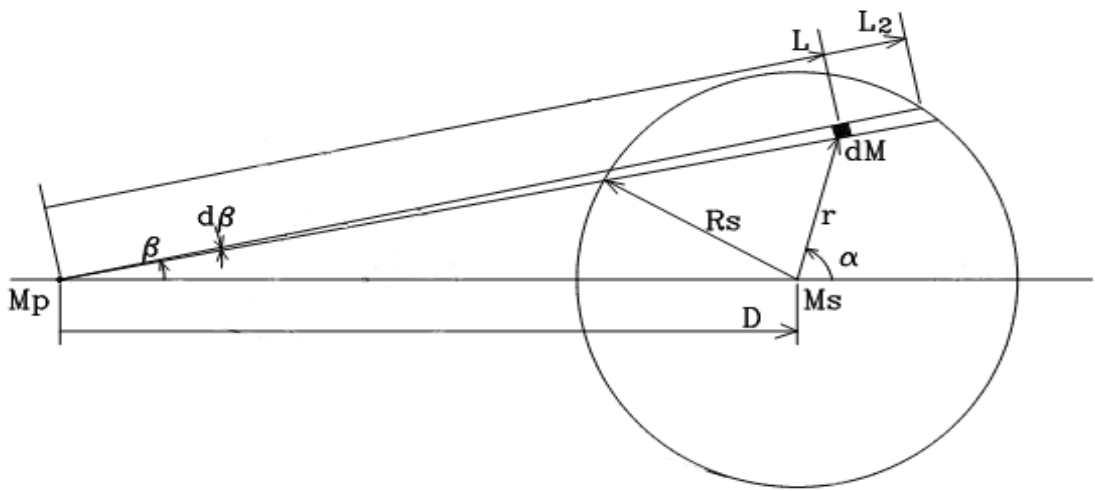

The variable D is the distance between the center of the Sun and Mp (identified at the center of the planet).
In the following calculations, each mass element dM is localized relative to its distance **r** to the center of the Sun and to the angle **α**, as shown in the figure above. Integration will be done on these two variables, **r** and **α**.

The angle b is given by   $\beta(D, r, \alpha) := atan\left(\dfrac{r \cdot sin(\alpha)}{D + r \cdot cos(\alpha)}\right)$

The distance between Mp and the point of entrance  into the Sun of the flux of gravitons, is

$$L2(D, r, \alpha) := \left[ D \cdot cos(\beta(D, r, \alpha)) + \sqrt{Rs^2 - (D \cdot sin(\beta(D, r, \alpha)))^2} \right]$$

The distance between Mp and the element dM is
$$L(D, r, \alpha) := \frac{\sin(\alpha)}{\sin(\beta(D, r, \alpha))} \cdot r$$

The traveled length inside the solar matter, by the flux of gravitons, is (L2 - L).

The gravitational "constant" is proportional to the flux of gravitons reaching *dM*. Its value is function of the absorption by the traversed matter and varies like *exp*(-K $\rho$ (L2 - L)), where $\rho$ is the density of the absorbant elements, and K their cross section per mass unit.

$$K := 3.27 \cdot 10^{-17} \quad m^2 \cdot kg^{-1}$$

Note : -We must use this value K in the following calculations, so that $\Delta\Omega c$ (definite further) takes a value equal to the experimental value (43 seconds of arc per century) for the advance of the perihelion of Mercury .-

In the calculations which follow, the effect of spatial distribution combined with the attenuation by the solar matter is taken into account for each annular mass element dM . Here dM is given by $\left(2 \cdot \pi \cdot \sin(\alpha) \cdot r^2 \cdot \rho \cdot dr \cdot d\alpha\right)$ .

**Attraction without effect of attenuation (law of Newton)**

$$gn(D) := \int_0^\pi \int_0^{Rs} \frac{G \cdot \left(2 \cdot \pi \cdot \sin(\alpha) \cdot r^2 \cdot \rho\right)}{L(D, r, \alpha)^2} \cdot \cos(\beta(D, r, \alpha)) \, dr \, d\alpha$$

The effect of attenuation for an element dM is $\qquad EC(D, r, \alpha) := \exp(-K \cdot \rho \cdot (L2(D, r, \alpha) - L(D, r, \alpha)))$

**Attraction with the effect of attenuation**

$$g(D) := \int_0^\pi \int_0^{Rs} \frac{(G \cdot EC(D, r, \alpha)) \cdot \left(2 \cdot \pi \cdot \sin(\alpha) \cdot r^2 \cdot \rho\right)}{L(D, r, \alpha)^2} \cdot \cos(\beta(D, r, \alpha)) \, dr \, d\alpha$$

The following expression has been modified to impose the effect of attenuation at the aphelion distance

With this formula, the computed value at the perihelion will be smaller

$$gpa(D) := \int_0^\pi \int_0^{Rs} \frac{(G \cdot EC(Da, r, \alpha)) \cdot \left(2 \cdot \pi \cdot \sin(\alpha) \cdot r^2 \cdot \rho\right)}{L(D, r, \alpha)^2} \cdot \cos(\beta(D, r, \alpha)) \, dr \, d\alpha$$

The following expression has been modified to impose the effect of attenuation at the perihelion distance

With this formula, the computed value at the aphelion will be greater

$$gap(D) := \int_0^\pi \int_0^{Rs} \frac{(G \cdot EC(Dp, r, \alpha)) \cdot \left(2 \cdot \pi \cdot \sin(\alpha) \cdot r^2 \cdot \rho\right)}{L(D, r, \alpha)^2} \cdot \cos(\beta(D, r, \alpha)) \, dr \, d\alpha$$

## c)  Comparative results: Corpuscular Theory vs Theory of  Einstein

The comparison is based on the evaluation of a variation of the gravitational "constant" between the perihelion and the aphelion. This variation is obtained by absorbing in G the corrective terms foreseen by each one of the two theories : Effect of attenuation for the Corpuscular Theory, Relativistic effect for General Relativity.

**Theory of Einstein**

Calculations which follow are deduced from the relation IV.95, p. 96, S. Mavridès, L'Univers relativiste, Masson, (1973). In this relation IV.95, the corrective term was absorbed in G. By withdrawing the value of G at the perihelion from the one at the aphelion, we obtain the variation of G between these two orbital positions. This variation dGn is given by

$$dGn := \frac{3 \cdot G^2 \cdot Ms}{c^2} \cdot Dga \cdot \left(1 - e^2\right) \cdot \left(\frac{1}{Dp^2} - \frac{1}{Da^2}\right) \qquad = \boxed{\frac{12 \cdot G^2 \cdot Ms}{c^2} \cdot \frac{e}{Dga \cdot \left(1 - e^2\right)}} \qquad dGn = 4.378 \bullet 10^{-18}$$

The multiplier $100 \cdot \dfrac{Tt}{T} \cdot \dfrac{360 \cdot 3600}{4 \cdot G \cdot e}$ that we apply to dGn gives us the advance of the perihelion in second of arc per century as formulated in the relation IV.116, p. 99, S. Mavridès, L'Univers relativiste, Masson, (1973).

$$\Delta\Omega e := 100 \cdot \frac{Tt}{T} \cdot \frac{360 \cdot 3600}{4 \cdot G \cdot e} \cdot dGn \qquad = \boxed{\frac{100 \cdot Tt}{T} \cdot \frac{3 \cdot G \cdot Ms}{c^2} \cdot \frac{360 \cdot 3600}{Dga \cdot \left(1 - e^2\right)}} \qquad \boxed{\Delta\Omega e = 42.95}$$

**Corpuscular theory**

We obtain the variation of the gravitational "constant" between the perihelion and the aphelion by making the difference for the same planetary distance between mean constants G, calculated with the effect of attenuation at the perihelion distance and at the aphelion distance. With the corrected law for the attenuation, we cannot bring back the expression as a central force. So, this way of calculating dGn gives the possibility of obtaining the difference of the average influence on G of the corrective term between the two orbital positions.

*Using the perihelion distance we have*

$$dpGn := \frac{\left(g(Dp)\right) \cdot Dp^2 - \left(gpa(Dp)\right) \cdot Dp^2}{Ms}$$

$$dpGn = 4.395 \bullet 10^{-18}$$

*Using the aphelion distance we have*

$$daGn := \frac{\left(gap(Da)\right) \cdot Da^2 - \left(g(Da)\right) \cdot Da^2}{Ms}$$

$$daGn = 4.386 \bullet 10^{-18}$$

In order to obtain the advance of the perihelion, in second of arc per century, we multiply **dpGn**, and **daGn** by $100 \cdot \dfrac{Tt}{T} \cdot \dfrac{360 \cdot 3600}{4 \cdot G \cdot e}$ such as we did for the theory of Einstein. It is logical to believe that if the corrective terms have the same effect on G, they will produce an identical advance of perihelion.

$$\Delta p\Omega c := 100 \cdot \frac{Tt}{T} \cdot \frac{360 \cdot 3600}{4 \cdot G \cdot e} \cdot dpGn \qquad\qquad \Delta a\Omega c := 100 \cdot \frac{Tt}{T} \cdot \frac{360 \cdot 3600}{4 \cdot G \cdot e} \cdot daGn$$

$$\Delta p\Omega c = 43.119 \qquad\qquad\qquad\qquad \Delta a\Omega c = 43.027$$

We will take the average, that is to say $\qquad \Delta\Omega c := \dfrac{\Delta p\Omega c + \Delta a\Omega c}{2} \qquad \boxed{\Delta\Omega c = 43.07}$

(In practice, several values of K are tested until obtaining 43 seconds of arc per centuries for Mercury)

**d) Advance of the perihelion of planets**

Using $K = 3.25 \cdot 10^{-17}$, we can start again the calculations after replacing the values of the parameters Dga, e and T, for each of the other planets (at the beginning of the section b). We obtain the following values for the four first planets:

| | **Mercury** | **Venus** | **Earth** | **Mars** |
|---|---|---|---|---|
| $\Delta\Omega e$ | 42.95 | 8.61 | 3.83 | 1.35 |
| $\Delta\Omega c$ | 43.07 | 8.63 | 3.84 | 1.35 |

Attenuation of gravitation through the matter is an excellent alternative to explain the advance of perihelion of planets. On another side, the same results with so dissimilar functions represent, in itself, a very interesting mathematical problem.